\documentclass[sigconf]{acmart}

\AtBeginDocument{%
  \providecommand\BibTeX{{%
    \normalfont B\kern-0.5em{\scshape i\kern-0.25em b}\kern-0.8em\TeX}}}

\copyrightyear{2025}
\acmYear{2025}
\setcopyright{acmcopyright}
\acmConference[Conference acronym 'XX]{Make sure to enter the correct
  conference title from your rights confirmation email}{June 03--05,
  2018}{Woodstock, NY}
\acmPrice{15.00}
\acmDOI{10.1145/3491102.3501940}
\acmISBN{978-1-4503-9157-3/22/04}

\usepackage{subcaption}
\usepackage{tabularx} 
\usepackage{booktabs}
\usepackage{geometry}
\usepackage{colortbl}
\usepackage{xcolor}
\usepackage{soul}
\usepackage{_macros}
\usepackage{caption}
\usepackage{subcaption}
\usepackage{array}
\usepackage{tikz}
\usepackage{hyperref}
\usepackage{hyperxmp}

\newcommand{\name}{\textbf{AnimAlte}}
\sethlcolor{blue!50!red!50!white}

\begin{document}

\title{AnimAlte: Designing AI-Infused Cartoon Videos to Improve Preschoolers' Language Learning with Family Engagement at Home}

\author{Shiya Tsang}
\orcid{0009-0008-4338-3570}
\affiliation{%
  \institution{Hong Kong University of Science and Technology (Guangzhou)}
  \city{Guangzhou}
  \country{China}
}
\email{szeng785@connect.hkust-gz.edu.cn}

\author{Ruiyao Miao}
\orcid{0009-0002-1848-4421}
\affiliation{%
  \institution{University of California, Los Angeles}
  \city{California}
  \country{United States}
}
\email{ruiyao0809@g.ucla.edu}

\author{Junren Xiao}
\orcid{0009-0006-5633-2268}
\affiliation{%
  \institution{Hong Kong University of Science and Technology (Guangzhou)}
  \city{Guangzhou}
  \country{China}
}
\email{jxiao767@connect.hkust-gz.edu.cn}

\author{Hui Xiong}
\authornote{Corresponding Author.}
\orcid{0000-0001-6016-6465}
\affiliation{%
  \institution{Hong Kong University of Science and Technology (Guangzhou)}
  \city{Guangzhou}
  \country{China}
}
\email{xionghui@hkust-gz.edu.cn}



\begin{abstract}

Cartoon videos have proven to be effective in learning vocabulary to preschool children. However, we have little knowledge about integrating AI into cartoon videos to provide systematic, multimodal vocabulary learning support. This late-breaking work present \name{}, an AI-powered cartoon video system that enables real-time Q\&A, vocabulary review, and contextual learning. Preliminary findings contextualized how families interact with \name{} to support vocabulary learning. Parents appreciated the system for its personalized, engaging experiences, fostering collaboration, and encouraging self-reflection on parenting. This study offers valuable design implications for informing future video systems to support vocabulary learning.

\end{abstract}




\begin{teaserfigure}
  \includegraphics[width=\textwidth]{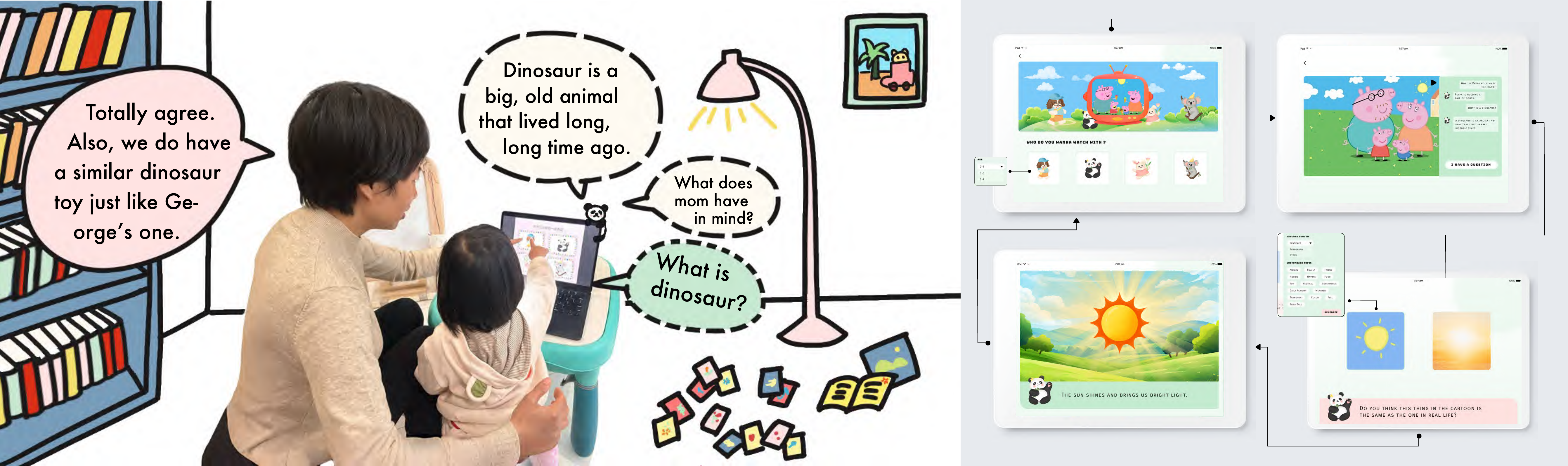}
  \vspace{-7mm}
  \caption{\name{} facilitates families to learning vocabulary through real-time question-answering, active review with questions and feedback,real-world associations by linking animated and real-life images, and contextual expansion with sentences or stories.}
  \Description{Enjoying the baseball game from the third-base
  seats. Ichiro Suzuki preparing to bat.}
  \label{fig:teaser}
\end{teaserfigure}


\maketitle

\section{Introduction}
Vocabulary plays a vital role in early childhood literacy development~\cite{song2015tracing,hoffman2014assessing}.
Research consistently shows that a well-developed vocabulary in preschool children might improve social interaction~\cite{erdemir2022vocabulary}, future academic success~\cite{cox2014children}, and reading comprehension~\cite{hadley2016examining}. 
An effective strategy for supporting preschool children's vocabulary learning is using cartoon videos, thanks to their unique style and expression~\cite{Vitasmoro2019/11,meng2020influence}.
Cartoon videos with repetitive input and contextual learning can engage children while improving vocabulary, pronunciation, and interest in language learning~\cite{perween2020impact,okunolaexploring,arifani2020cartoon}.
However, most cartoon videos rely on passive input, offering limited opportunities for interaction with either the content or caregivers, which might reduce their effectiveness for language learning~\cite{10.1145/3078072.3079717,meng2020influence}.
Learners also tend to forget information from video-based learning since it isn't systematically processed, retained, or reviewed~\cite{goodianti2007learners}. 
Additionally, studies suggest young children may struggle to distinguish between real and fantastical events, hindering their ability to apply the language learned from cartoon in different context of real life~\cite{Beege2024,Prosic-Santovac2017,Vitasmoro2019/11,li2015can}.

In the field of Human-Computer Interaction (HCI), AI-infused video learning has emerged as an innovative approach within the context of education and learning~\cite{10.1145/3544548.3581153,10.1145/3544549.3585804,10.1145/3442381.3449808,10.1145/3613904.3642587}. 
For example, Xu et al. showed that incorporating a conversational agent into video programs, which engages children with questions and feedback, enhances active learning during video watching~\cite{10.1145/3491102.3502050}.
However, few studies have explored how AI conversational agents can understand video frames to provide offer personalized, interactive feedback that enhances preschoolers' language learning.
In the non-video learning context, research shows that large language models (LLMs) improve vocabulary learning by generating personalized, contextually relevant content~\cite{leong2024putting}. 
Additionally, AI-generated images may aid in learning target words through reading and retelling a story~\cite{chen2024retassist}. 
These studies highlight the potential of AI to enhance preschool language learning, suggesting an opportunity to integrate AI into cartoon videos for a more systematic and multimodal approach.

To this end, we proposed an AI-infused cartoon video system, \name{}, designed to support preschool children's vocabulary learning~(\autoref{fig:teaser}). 
This system features four phases for supporting vocabulary learning with family engagement through real-time question-answering during cartoon videos, active review with questions and feedback, real-world associations by linking animated and real-life images, and contextual expansion with sentences or stories tailored to topics and age groups. 
In our late-breaking work, we conducted a user study involving 5 pairs of parents and children to explore how families interact with \name{} for vocabulary learning in home settings (\textbf{RQ1}) and how they perceive cartoon-based vocabulary learning using \name{} (\textbf{RQ2}).
The preliminary findings helped contextualize the process of how families interact with \name{} to learn vocabulary. The data also illustrated how parents appreciated \name{} for providing their personalized and engaging learning experiences, helping them reflect on their parenting and fostering their collaboration.
\section{FORMATIVE STUDY}
To inform the design of \name{}, we first conducted semi-structured interviews with parents to understand current practices and challenges in using cartoon videos to support preschoolers' language development. 
Next, we interviewed experts to discuss challenges identified from the parent interviews and explore the potential role of AI in enhancing language learning for preschoolers.

\subsection{Procedure and Analysis}
First, we recruited six families (mothers and their children aged 2–5) through a previously partnered kindergarten in [the city of Anonymity], with demographics detailed in APPENDIX \autoref{tab:family1}.
All families had experience using cartoon videos for language learning. We conducted 30-minute interviews with each parent to explore their practices and challenges in using cartoons for vocabulary development. Additionally, we held one-hour interviews with three child language specialists and a speech-language pathologist, all experienced in early childhood education. We introduced AI technologies such as visual language models (VLMs), large language models (LLMs), and image-based generative AI, emphasizing their natural language capabilities, and invited insights on their potential for video-based learning.

All sessions were recorded, transcribed, and analyzed using open coding and affinity diagramming~\cite{braun2006using}. The study was approved by the research ethics board, with pseudonyms used for participants. Parents received 100 RMB (~13 USD), and experts were compensated at their standard hourly rate.

\subsection{Findings from the Interviews}
\subsubsection{CH1: Challenges in Interactive Feedback in Cartoon-Based Vocabulary Learning} 
Most parents mentioned that their children lack interactive feedback while learning vocabulary through cartoon videos. P6 mentioned that the lack of interaction with children may limit children's motivation to express curiosity: ``\textit{My child is naturally curious. He asked about the meaning of taxi colors while watching cartoons. If I'm distracted by my phone, he loses interest in continuing to ask}''. Our experts highlighted that the lack of interactive feedback in vocabulary learning through cartoons might result in poor outcomes, as ``\textit{language is a two-way tool, while cartoons offer one-way interaction, leaving children passive and unable to correct mistakes or deepen their understanding(E1)}''.

Experts emphasized that parental involvement, through interactive feedback, can support preschool children's vocabulary learning and strengthen the parent-child relationship during cartoons. 
However, some parents mentioned struggling to understand and respond to their children's questions, leading to communication barriers. P10 noted that when children asked questions about the plot of the animation,  it might be difficult for her to answer if she had not watched the cartoon, thus reducing the interaction with the children. Also, P12 mentioned that her child's questions about Vienna were beyond her knowledge.
Further, many parents struggled to adjust their communication to their child's developmental stage, e.g., P4 may need to speak more slowly and use simple, cute words to help their child understand, as the child might not fully grasp it otherwise.

\subsubsection{CH2: Challenges in Vocabulary Review through Cartoon Videos}
Our parents indicated that using cartoon videos to help preschool children learn new vocabulary might lack support for repetitive review. 
For instance, P8 suggested that cartoons may not offer enough memory support for recalling newly learned vocabulary: ``\textit{The last time we watched Frozen, my child learned the word `brave', but after the cartoon ended, P7 didn't seem to mention the word again}''.
To address this, P4 replayed specific cartoon segments to help her children review vocabulary. P8 wrote vocabulary or sentences on the walls to read with her child, while E1 took screenshots of cartoon scenes for effective review.
Some parents complained that when their child asked too many questions or discussed too much vocabulary during an episode, the review process became time-consuming and tedious, burdening them with organizing and revisiting the material.

\subsubsection{CH3: Challenges in Bridging the Connection Between the Virtual and Real Worlds}
Most parents noted that characters and objects in cartoon videos often seem disconnected from the real world, e.h., P8 worried that anthropomorphized characters in cartoons might confuse her cognition, so she used real photos of objects, like cards with apples, to help the child learn these words.
E1 explained that disconnection from the real world could hinder learners' ability to relate new words to real-life contexts, making it harder to internalize and apply them meaningfully.
E1, E3, and E4 stressed the importance of comparing animated items with real objects to aid children's vocabulary understanding. E1 helped children recognize cartoon items by buying their real-life counterparts, while E4 connected the cartoon sun to real sunsets and noons, helping children identify the sun in various images and link the word to their everyday experiences.

\subsubsection{CH4: Challenges in Contextualizing Vocabulary Beyond Cartoon Videos}
Several parents mentioned that learning vocabulary through watching cartoon videos sometimes lacks application and extension in various contexts.
For example, P4 observed that after watching cartoons, their child could name objects and animals but struggled to talk about them in different settings. P6 noted that learning vocabulary and events through cartoons might lack sentence expansion or story extension, limiting the child's broader thinking: ``\textit{My child learned the name `apple', but I felt that wasn't enough. I wanted to expand by sharing the story of Newton and the apple}''.
E1 emphasized the importance of expanding sentences and stories, noting that learners can not only recognize an onion's appearance but also understand its taste and uses through stories and sentences, deepening their vocabulary and gradually forming a semantic network.

\begin{figure*}[ht]
    \centering  \includegraphics[width=1\linewidth]{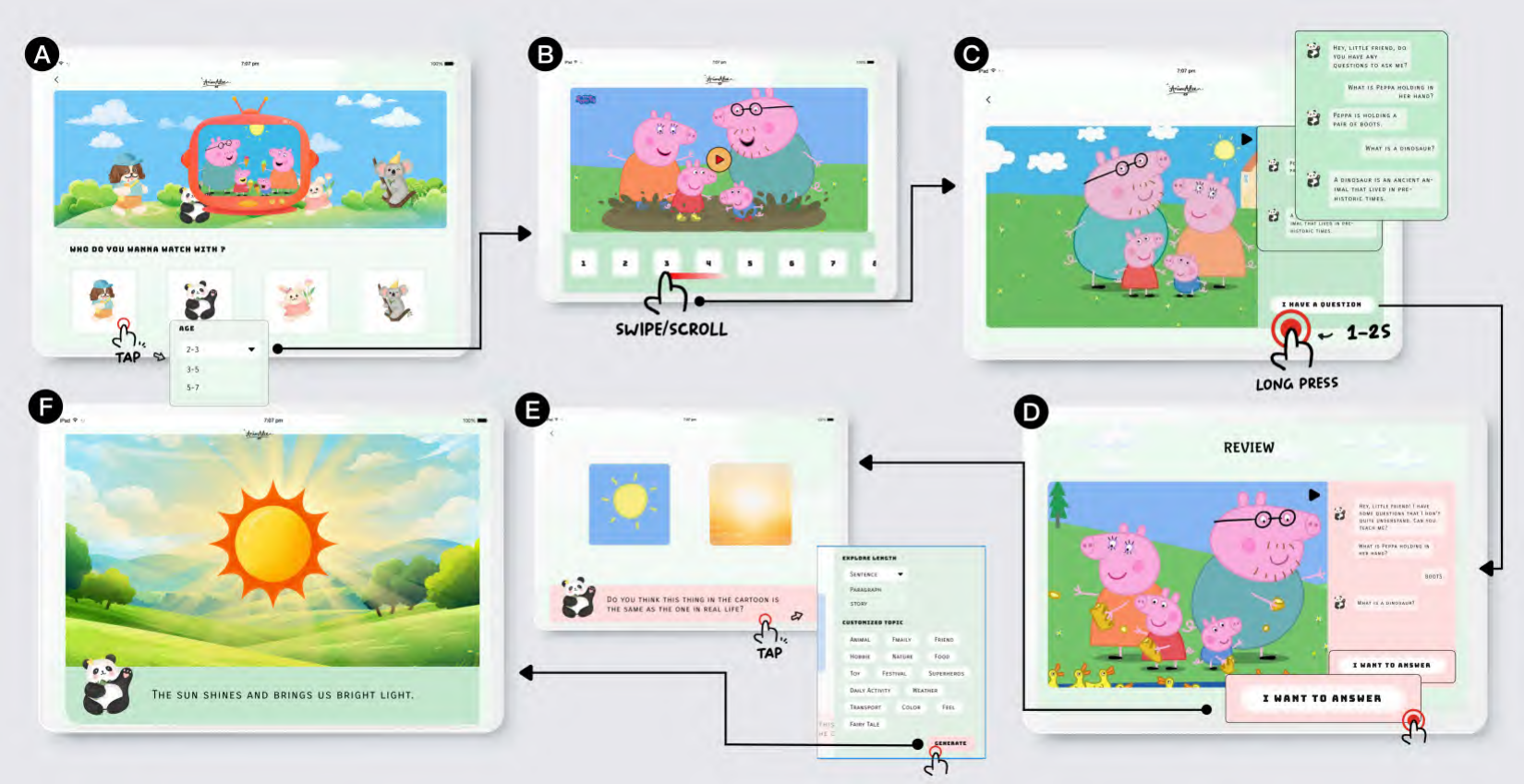}
    \caption{User interface flow of \name{} (A), episode navigation (B), cartoon watching and question answering (C), review (D), real-world association (E), and contextual expansion (F).}
    \label{fig:ui_interface}\vspace{-0.5cm}
\end{figure*}
\section{\name{} SYSTEM}
Informed by the formative study, we designed and developed \name, an AI-infused cartoon video system for promoting preschoolers' language learning. In this section, we discuss our core features, usage scenario, and system implementation.
\subsection{Core Features}
The core features of \name{} were informed by our formative study, existing literature, and insights from experts, culminating in the identification of four key phases: 
The \textbf{Real-time Question-Answering} phase allows families to pause the video and interact with a conversational agent, which answers questions about the current frame to help children learn vocabulary tied to the cartoon's objects and events.
The \textbf{Active Vocabulary Review} phase features the agent, acting as a teachable agent~\cite{matsuda2020effect}, curiously asking children the same questions, presenting relevant screenshots, and providing positive feedback to reinforce newly learned vocabulary.
The \textbf{Real-World Association} phase pairs real-world images with their animated versions to help children connect the object to its real-world counterpart.
The \textbf{Contextual Expansion} phase creates personalized stories and sentences based on family preferences and children's ages~\cite{leong2024putting}. Drawing from previous work~\cite{chen2024retassist}, it also generates related images to enhance word comprehension and recall in context.

\subsection{\name{} Usage Scenario}
\autoref{fig:ui_interface} illustrates the overall user flow of \name{} which is an AI-infused cartoon video system for vocabulary learning. Here we present a typical usage scenario of \name{}: 
Oli is a 3-year-old child who loves watching the cartoon \textit{Peppa Pig}. 
Her mother, Ira, wanted her to learn new vocabulary through the cartoon video. 
One evening, they opened \name{} on a tablet, and Ira encouraged Oli to choose a favorite cartoon character to watch with her (\autoref{fig:ui_interface} (A)). 
Ira assisted Oli in selecting a panda character and the appropriate age group, ensuring the agent could communicate effectively with Oli. 
Then, Ira selected an episode of \textit{Peppa Pig} for them to watch together (\autoref{fig:ui_interface} (B)). 
While watching, Ira paused the video, pointed to the noodles in the cartoon, and asked Oli, ``What is Peppa eating?'', and Oli shook her head.
Ira clicked ``I have a question'' in the interface and asked again. The agent replied, ``Peppa is eating noodles''~(see \autoref{fig:ui_interface} (C)).
Further, Ira encouraged Oli to ask the conversational agent if there was something she didn't understand. 

After watching the video, the conversation agent transformed into a curious teachable agent by showing Oli screenshots from the cartoon and asking her to review the vocabulary she just learned. 
For example, the conversational agent asked, ``What is Peppa eating?''.  Oli could click the ``I want to answer'' button, and the agent would give feedback based on her response (see \autoref{fig:ui_interface} (D)).

In the Contextual Mapping phase, Oli and her mom saw two images: one of noodles from Peppa Pig on the left, and a real-life image on the right. The conversational agent prompted a discussion question, asking if the two images were similar. Oli and her mom then discussed it, with her mom adding, ``Have you seen the noodles we eat at home?'' (see \autoref{fig:ui_interface} (E)).

Finally, they found that the system expanded vocabulary into sentences or stories and generated corresponding images. For example, Ira customized the sentence under the topic ``FOOD'': ``Noodles are made from flour, and when cooked, they become soft and pair perfectly with various delicious soups'', expanding from the word ``noodles'' (see \autoref{fig:ui_interface} (F)).

\begin{figure*}[h]
    \centering
    \includegraphics[width=0.9\linewidth]{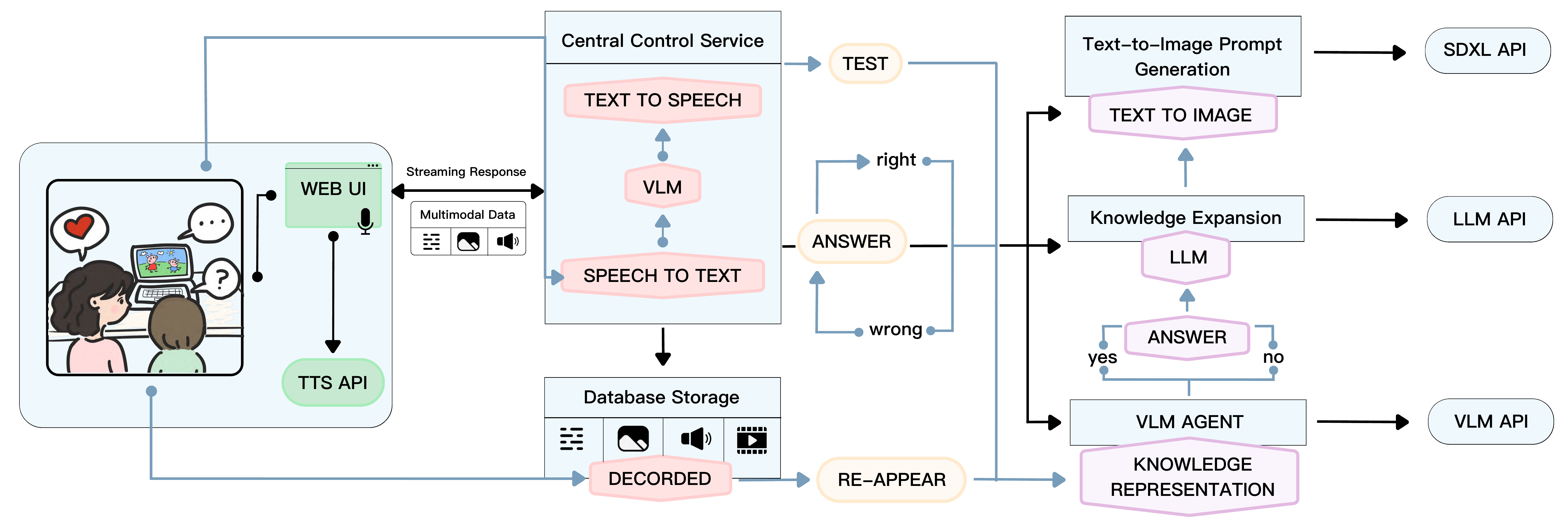}
    \caption{: Overview of \name{} system architecture.}
    \label{fig:enter-label}\vspace{-0.5cm}
\end{figure*}

\subsection{ \name{} System Implementation}
\label{subsec:system-implementation}
\autoref{fig:enter-label} shows our system architecture. 
The front-end application of \name{} is built with \textit{Vue 3}~\cite{vue3}, a JavaScript-based framework that enables cross-platform compatibility across devices. The client communicates with the server via \textit{REST API}. When interacting with \name{}, user audio messages and video frames are uploaded to the backend as input. User audio messages are processed using \textit{iFly's TTS}
\footnote{iFly's TTS, https://global.xfyun.cn/products/text-to-speech} model to generate corresponding voice messages. Also, we have implemented speech-to-text function using the Voice Dictation (Streaming Version) service from \textit{iFlytek}\footnote{iFLYTEK Speech-to-Text, https://global.xfyun.cn/products/speech-to-text}. 

The back-end of \name{} is built with FastAPI~\cite{fastapi}, which facilitates the processing of VLMs and LLMs workflows, AI image generation, and data management. 
The generative pipelines incorporate \textit{CogVLM2}\footnote{CogVLM2, https://github.com/THUDM/CogVLM} to analyze video frames during Real-time Q\&A phase and Active Review phase to provide corresponding feedback. 
In the Real-World Association phase, the conversation data between the agent and families from the first phase was summarized into themes using \textit{Doubao LLM}\footnote{Doubao AI, https://team.doubao.com/zh/}, then transformed into image prompts with realistic styles by another \textit{Doubao LLM}, which were input into Stable Diffusion to generate images.
In the Contextual Expansion phase, we used \textit{Doubao LLM}\footnote{Doubao AI, https://team.doubao.com/zh/} to generate contextual sentences and stories, which are then summarized into prompts for the Stable Diffusion model to generate images (\textit{stabilityai/stable-diffusion-xl-base-1.0}\footnote{Stable Diffusion XL, https://huggingface.co/docs/diffusers/using-diffusers/sdxl}).
We iteratively co-designed AI prompts with language experts.
The system uses MongoDB to store and manage data, including user queries, interaction logs, and AI-generated content (e.g., video frames, dialogue records, and AI-generated sentences or stories).
\vspace*{-0.2cm}\section{Pilot Study and Evaluation}
We conducted a remote pilot study to explore two key questions: how families engage with AI-infused cartoon videos to learn vocabulary in home settings (\textbf{RQ1}), and how they perceive cartoon-based vocabulary learning with \name{} (\textbf{RQ2}).

\subsection{Participants}
The research team recruited family groups by distributing digital flyers on a social media platform, describing the pilot study as an activity to enhance preschoolers' language learning through a digital cartoon video system. Five families participated, comprising 10 individuals: 5 children (ages 3–5), 4 mothers, and 1 father. Participant demographics are detailed in APPENDIX \autoref{tab:familystudy}, with all families sharing the same native language to ensure cultural and linguistic consistency. The study was approved by the institutional ethics board, and pseudonyms were used to protect anonymity. Each family received 100 RMB as compensation.


\subsection{Study Procedure and Data Analysis}
Each hour-long user study session was conducted remotely via Tencent Meeting. Families accessed \name{} on their personal computers or tablets through a web browser, sharing their screens with researchers and enabling cameras when possible. Before the study, We held a 20-minute online session to introduce the activity, using slides to demonstrate \name{}'s interface and features. Families were encouraged to explore the system freely, with parents asking 2-3 questions based on their understanding of the cartoon videos and motivating their children to ask questions as well. After the introductory session, each family pair used \name{} to watch an episode of \textit{Peppa Pig}. This cartoon was chosen for its suitable length, everyday language, rich visual cues, and relevance to children's lives, aligning with language development goals. Parents utilized \name{}'s interactive features during the four phases (Section 3.2) to support collaborative vocabulary learning. Researchers acted only as technical support, without intervening in the session. After using \name{}, we conducted 20-minute semi-structured interviews with each parent to explore our research questions. All sessions were recorded, totaling 347 minutes of audio and 216 minutes of screen recordings, which were transcribed for thematic analysis~\cite{braun2006using}.
\section{FINDINGS}
\subsection{RQ1: How Families Interacted with \name{} to Learn Vocabulary in Home Settings}
In this pilot study, we contextualized the process of families interact with \name{} to learn vocabulary: In \textbf{Real-Time Q\&A} phase, several parents utilized real-time Q\&A interactions to expand their knowledge of vocabulary-related topics, e.g, C1 mentioned that they enhanced children's vocabulary by asking for additional words related to those they had already learned: ``\textit{We saw Peppa eating spaghetti. I asked the AI, `Why is the pasta yellow?' and it explained that it's made from wheat, helping me learn why it's yellow}''. In \textbf{Active Vocabulary Review} phase, interactions between parents and children often followed a turn-taking pattern. When C7 hesitated to review certain vocabulary, the parent acted as a ``role model'', answering the conversational agent's questions and demonstrating correct responses. This encouraged C8 to motivate the child to answer the next question, gradually fostering more active participation in the review process. In \textbf{Real-World Association} phase, C4 utilized real-life object images from \name{} to encourage children to discuss whether they have these items at home, connecting cartoon vocabulary to their living environment. Similarly, C6 used real-life images to spark multisensory discussions, asking questions like, ``What does this onion smell like? Do you remember the sound when Dad was cutting it that day?''. In \textbf{Contextual Expansion} phase, C10 customized sentences based on the vocabulary children learn, focusing on topics they frequently encounter in daily life, e.g., food-related topics helped children understand the various uses of different foods. Meanwhile, the AI-generated stories for C8 and C7 adopted a more interactive approach, encouraging them to take on roles within the narrative and retell the story.

\subsection{RQ2: How They Perceived Cartoon-based Vocabulary Learning with \name{}}
In our study, parents valued the system for its ability to provide personalized and engaging learning experiences for preschool children. 
C10 appreciated that AI can offer tailored responses based on the child's vocabulary level: ``\textit{He only knew basic words like `apple' and `banana', and I found that the AI responded with age-appropriate, simple sentences. The voice was also cute, something I couldn't replicate}".
C8 also noted that AI-generated images might enhance children's engagement during sentence expansion: ``\textit{For 3-year-olds with limited interest in text, highly relevant visuals combined with expanded sentences can better capture their attention, and I noticed she really enjoyed these images}.

Some parents appreciated the system for encouraging self-reflection on their parenting. For example, C4 noted how the AI patiently answered C3's random questions, saying, ``\textit{I asked how many pigs were in the animated scene... and then if I could provide a formula. The AI answered even absurd queries, and C3 enjoyed chatting with it. I didn't respond to these questions and began to wonder if I might be a bit too impatient}''.

Finally, our parents reported that \name{} can foster their collaboration during learning vocabulary. C6 stated that the real-time Q\&A might as a great medium to spark more questions, encouraging deeper communication beyond vocabulary: ``\textit{I inquired about what George was holding, and the AI responded with a dinosaur. I then asked my child if he knew how dinosaurs became extinct. After he shook his head, I encouraged him to ask the AI}''. Also, C2 perceived the teachable agent as a bridge during vocabulary review, encouraging the child to seek help: ``\textit{The AI prompted the child to ask about words they'd learned. If the child didn't understand, they'd come to me for an explanation, which I encouraged them to share with the AI}''.
\section{Design Implications and Future Work}
Here, we outline and explore a set of design implications to guide future HCI research:

\textbf{Implication 1: Integrating Personalized Question-and-Answer Interactions for Vocabulary Learning During Cartoon Watching}
Prior research has shown that integrating conversational agents into narrative science programming to promote active learning in children through questions and responsive feedback~\cite{10.1145/3491102.3502050}.
Extending the line of research, our system employs visual language models tailored to specific child age groups, helping families answer questions about cartoon video frames using age-appropriate vocabulary and expressions.
Future work could explore incorporating more personalized vocabulary assessment mechanisms into cartoon-based learning by enabling conversational agents to actively ask families questions based on screen frames. For instance, sensor devices (e.g., wearable devices, home environment sensors) could be used to collect children's daily language environments and identify priority vocabulary for learning~\cite{lee2024open}. Meanwhile, deep learning models could detect key scene elements in cartoons (e.g., characters, animals, actions) and dynamically generate personalized questions by integrating data from the children's language environments.

\textbf{Implication 2: Supporting Multimodal Interactions to Strengthen Real-World Associations}
Cartoons, popular among preschoolers, often feature fantastical events that defy physics. Research shows young children might struggle to distinguish between real and fantastical events~\cite{li2015can}.
To develop the ability to distinguish fantasy from reality, our system provides real-life images and cartoon screenshots for comparison to strengthen this distinction.
Future work could consider incorporating more multi-sensory interactions to enhance the ability to distinguish between reality and the virtual world. For example, animal sound effects could be incorporated to help users recognize real-world animals, or AI-generated interactive 3D models of objects could allow users to explore the details of these items~\cite{ma2024research}. Further, we also encourage considering the use of AI to generate 3D worlds from everyday images in the future, allowing children to engage with dynamic, immersive content~\cite{worldlabs}.

\textbf{Limitations and Future Work} 
In this section, we discuss several limitations of the current study and outline potential directions for future research:
The small sample size may limit result generalizability. Future work will involve recruiting more families for a long-term field study to gain deeper insights into how parents and children use \name{} in daily life. Second, the children in our study only watched \textit{Peppa Pig}. While they showed strong interest, future work should include more cartoons for different age groups to better support vocabulary learning and cater to a wider range of child participants.
Another key consideration is the safe use of generative AI. Although no inappropriate images appeared during the contextual expansion phase, parents noted minor issues with counterfactual elements in the AI-generated images. Future work could fine-tune generative models for children's vocabulary learning to ensure contextually accurate and appropriate images by training on datasets focused on children's cognitive development and vocabulary.
While visual language models can capture cartoon details and provide interactive feedback, their understanding of context remains limited. Future work could integrate live multimodal LLMs~\cite{google_ai_studio} and memory functions to enhance real-time contextual comprehension, enabling more comprehensive answers.

\bibliographystyle{ACM-Reference-Format}
\bibliography{reference.bib}

\newpage
\appendix
\section{Appendix}

\definecolor{customcolor}{HTML}{FEF9F2} 
\begin{table*}[tb]
\centering
\tabcolsep=0.5cm
\begin{tabular}{cccccc}\hline
\textbf{ID} & \textbf{Family Number} & \textbf{Role} & \textbf{Gender} & \textbf{Age} & \textbf{Educational Level} \\\hline
01 & F1 & Daughter & F & 2  & K1         \\\rowcolor{customcolor}
02 & F1 & Mother   & F & 28 & Bachelor’s \\
03 & F2 & Son      & M & 3  & K1         \\\rowcolor{customcolor}
04 & F2 & Mother   & F & 30 & Bachelor’s \\
05 & F3 & Son      & M & 3  & K1         \\\rowcolor{customcolor}
06 & F3 & Mother   & F & 32 & —          \\
07 & F4 & Daughter & F & 4  & K2         \\\rowcolor{customcolor}
08 & F4 & Mother   & F & 34 & Bachelor’s \\
09 & F5 & Son      & M & 4  & K2         \\\rowcolor{customcolor}
10 & F5 & Mother   & F & 32 & Master’s   \\
11 & F6 & Daughter & F & 5  & K2         \\\rowcolor{customcolor}
12 & F6 & Mother   & F & 36 & Bachelor’s\\\hline
\end{tabular}
\caption{Demographics of Participant Families from the Formative Study: The Family Number refers to the specific family to which the participant is assigned.}
\label{tab:family1}
\end{table*}

\definecolor{customcolor}{HTML}{FEF9F2} 
\begin{table*}[tb]
\vspace{-0.5cm}
\centering
\tabcolsep=0.5cm
\begin{tabular}{cccccc}
\hline
\textbf{ID} & \textbf{Family Number} & \textbf{Role} & \textbf{Gender} & \textbf{Age} & \textbf{Educational Level} \\\hline
01 & F1 & Son      & M & 4  & K2         \\\rowcolor{customcolor}
02 & F1 & Mother   & F & 28 & Bachelor’s \\
03 & F2 & Son      & M & 5  & K3         \\\rowcolor{customcolor}
04 & F2 & Mother   & F & 49 & Bachelor’s \\
05 & F3 & Son      & M & 5  & K3         \\\rowcolor{customcolor}
06 & F3 & Father   & M & 40 & Doctorate (PhD) \\
07 & F4 & Daughter & F & 3  & K1         \\\rowcolor{customcolor}
08 & F4 & Mother   & M & 47 & Bachelor’s \\
09 & F5 & Son      & M & 3  & K1         \\\rowcolor{customcolor}
10 & F5 & Mother   & M & 29 & Bachelor’s \\\hline
\end{tabular}
\caption{Demographics of Participant Families from Pilot Study: The Family Number refers to the specific family to which each participant belongs.}
\label{tab:familystudy}
\end{table*}


\end{document}